\documentclass[preprint,showpacs]{revtex4}    

\usepackage{graphicx}

\begin{document}


\title{First-principles Study of Physisorption of Nucleic Acid Bases
on Small-Diameter Carbon Nanotube}

\author{S. Gowtham$^1$}
\author{Ralph H. Scheicher$^{1,2}$}
\author{Ravindra Pandey$^1$}\email[Corresponding authors. E-mail: ]{rhs@mtu.edu, pandey@mtu.edu}
\author{Shashi P. Karna$^3$}
\author{Rajeev Ahuja$^{2,4}$}

\affiliation{$^1$Department of Physics and Multi-Scale Technologies
Institute, Michigan Technological University, Houghton, Michigan
49931, USA}

\affiliation{$^2$Condensed Matter Theory Group, Department of
Physics, Box 530, Uppsala University, S-751 21 Uppsala, Sweden}

\affiliation{$^3$US Army Research Laboratory, Weapons and Materials
Research Directorate, ATTN: AMSRD-ARL-WM; Aberdeen Proving Ground,
Maryland 21005-5069, USA}

\affiliation{$^4$Applied Materials Physics, Department of Materials
and Engineering, Royal Institute of Technology (KTH), S-100 44
Stockholm, Sweden}

\date{\today}

\begin{abstract}
We report the results of our {\it first-principles} study based on
density functional theory on the interaction of the nucleic acid
base molecules adenine (A), cytosine (C), guanine (G), thymine (T),
and uracil (U), with a single-walled carbon nanotube (CNT).
Specifically, the focus is on the physisorption of base molecules on
the outer wall of a (5,0) metallic CNT possessing one of the
smallest diameters possible. Compared to CNTs with large diameters,
the physisorption energy is found to be reduced in the
high-curvature case. The base molecules exhibit significantly
different interaction strengths, and the calculated binding energies
follow the hierarchy G $>$ A $>$ T $>$ C $>$ U, which appears to be
independent of the tube curvature. The stabilizing factor in the
interaction between the base molecule and CNT is dominated by the
molecular polarizability that allows a weakly attractive dispersion
force to be induced between them. The present study provides an
improved understanding of the role of the base sequence in
deoxyribonucleic acid (DNA) or ribonucleic acid (RNA) on their
interactions with carbon nanotubes of varying diameters.
\end{abstract}


\maketitle

\section{Introduction}

There has been a steady increase in interest over the past four
years in the non-covalent interaction of DNA with carbon nanotubes
(CNTs). This hybrid system at the junction of the biological regime
and the nanomaterials world possesses features which makes it very
attractive for a wide range of applications. Initially, the focus
rested on a new way to disperse CNT bundles in aqueous solution
\cite{Nakashima:2003} and to create a more efficient method to
separate CNTs according to their electronic properties
\cite{Zheng:2003a, Zheng:2003b}. More recently, interest has shifted
towards applications aimed at electronic sensing of various odors
\cite{Johnson:2005}, or probing conformational changes in DNA
\textit{in vivo} triggered by change in the surrounding ionic
concentration \cite{Strano:2006a}. It has been shown that
hybridization between complementary strands of DNA could be detected
on the surface of a CNT as well \cite{Star:2006,Strano:2006b}.
Finally, DNA may not only interact with the outer surface of CNTs,
but can be also be inserted inside CNT \cite{Okada:2006}, which may
allow for further potential applications of this particular nano-bio
system.

The details of the interaction of DNA with CNTs have not yet been
fully understood, though it is generally assumed to be mediated by
the $\pi$-electron networks of the base parts of DNA and the
graphene-like surface of CNTs
\cite{Zheng:2003a,Ortmann:2005,Seifert:2007}. It is therefore
desirable to obtain a better understanding of the binding mechanism,
and the relative strength of base-CNT binding as it is indicated
experimentally from sequence-dependent interactions of DNA with CNTs
\cite{Zheng:2003b,Johnson:2005}. Here, we present the results of our
{\it first-principles} study of the interaction of nucleic acid
bases with a (5,0) metallic CNT \cite{Li:2001, Machon:2002,
Liu:2002, Cabria:2003} as a significant step towards an
understanding of the fundamental physics and the mechanism of this
sequence-dependent interaction of ssDNA with CNTs.

In the present study, we have considered all five nucleic acid bases
of DNA and RNA: the two purine bases - adenine (A) and guanine (G),
and the three pyrimidine bases - cytosine (C), thymine (T), and
uracil (U). Our specific interest is to assess the subtle
differences in the adsorption strength of these nucleic acid bases
on a CNT with a very small diameter. Recently, we investigated the
interaction of DNA and RNA base sequences with a planar graphene
sheet \cite{Gowtham:2007}. The present effort is complementary to
the previous study, since the graphene sheet can be seen as a model
system for CNTs with a diameter much larger than the dimensions of
the bases, and hence a negligible curvature. Comparison of the two
sets of results allows us to determine the influence that curvature
has on the interaction of DNA and RNA with CNTs.

\section{Computational Method}

We employed a supercell approach in all our calculations. The unit
cell of a (5,0) single-walled carbon nanotube, consisting of a ring
of 20 carbon atoms with a diameter of 3.92 \AA\ was repeated three
times along the tube axis. In the direction perpendicular to the
tube axis, a distance of at least 15 \AA\ was kept between repeated
units to avoid interactions between adjacent CNTs.

The base molecules were terminated with a methyl group where the
bond to the sugar ring had been cut in order to generate an
electronic environment in the nucleic acid base more closely
resembling the situation in DNA and RNA rather than that of just
individual isolated bases by themselves. This has the additional
benefit of introducing a small magnitude of steric hindrance due to
the methyl group, quite similar to the case in which a nucleic acid
base with attached sugar and phosphate group would interact with the
surface of the CNT.

Calculations were performed using the plane-wave pseudopotential
approach within the local density approximation (LDA) \cite{LDA} of
density functional theory (DFT) \cite{DFT}, as implemented in the
Vienna Ab-initio Simulation Package ({\sc vasp}) \cite{VASP}.
Results were found to converge for a cutoff energy of 850 eV. We
used a $1\times1\times3$ Monkhorst-Pack grid \cite{Monkhorst:1976}
for $k$-point sampling of the Brillouin zone. In our previous study
on graphene \cite{Gowtham:2007}, $1\times1\times 1$ was found to
yield virtually identical results as that of a $3\times3\times1$
Monkhorst-Pack grid.

It has been reported \cite{Simeoni:2005, Tournus:2005} that the LDA
approximation appears to give a reliable description of dispersive
interactions, unlike the generalized gradient approximation (GGA)
\cite{GGA} for which binding is basically non-existent for van der
Waals bound systems. In a study of the adsorption of the base
molecule A on graphite \cite{Ortmann:2005} using LDA and a modified
version of the London dispersion formula \cite{London:1930} for van
der Waals (vdW) interactions in combination with GGA, it was found
that LDA, while underbinding the system, does in fact yield a
potential energy surface which is almost indistinguishable in its
structure from the one obtained via the GGA+vdW approach (cf. Figs.
1a and 1b of Ref. \cite{Ortmann:2005}). Furthermore, LDA yields
almost the same equilibrium distance of A to graphene as GGA+vdW.

\begin{figure}[ht]
 \begin{center}
    \includegraphics[scale=0.50, angle=-90.]{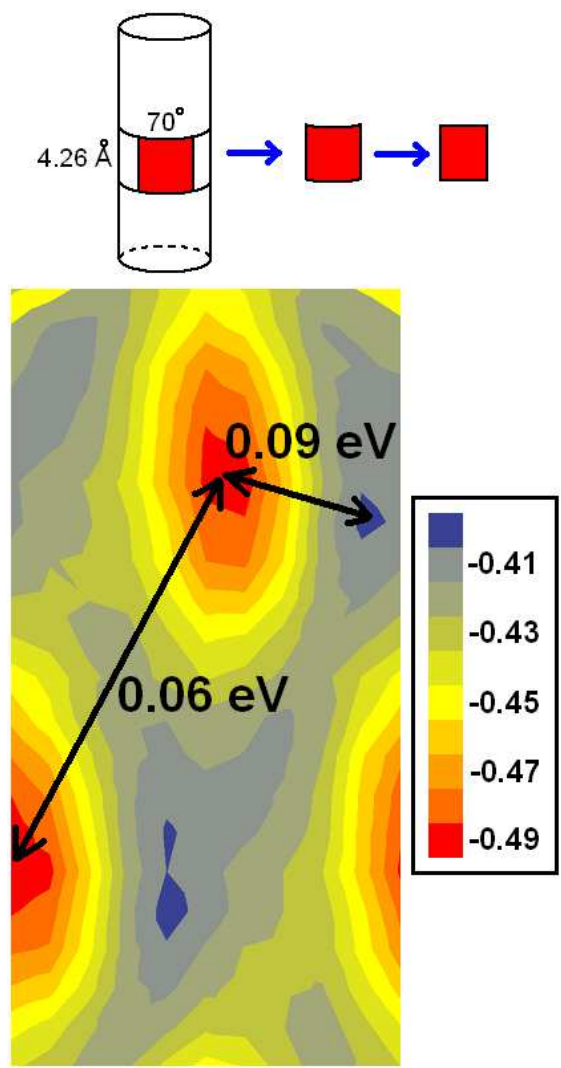}
    \caption{Potential energy surface (PES) plot (in eV) for guanine with CNT.
    Qualitatively similar PES plots were obtained for the other four base molecules.
    The scanning area is indicated by a red rectangle. The energy range between peak and
    valley is approximately 0.09 eV, while the energy barriers between adjacent global
    minima is only 0.06 eV.}
    \label{PES}
  \end{center}
\end{figure}

Following a similar procedure employed in our previous study with
graphene \cite{Gowtham:2007}, we started by carrying out the
optimization process as follows: (i) an initial force relaxation
calculation step to determine the preferred orientation and optimum
height of the planar base molecule relative to the surface of the
CNT; (ii) a curved slice of the potential energy surface was then
explored by translating the relaxed base molecules parallel to the
CNT surface covering a surface area 4.26 \AA\ in height, 70$^\circ$
in width (Fig. \ref{PES}) and containing a mesh of 230 scan points
(the separation between base molecule and the surface of the CNT was
held fixed at the optimum height determined in the previous step);
(iii) it was subsequently followed by a 360$^\circ$ rotation of the
base molecules in steps of 5$^\circ$ to probe the energy dependence
on the orientation of the base molecules with respect to the
underlying CNT surface; (iv) finally, a full optimization was
performed in which all atoms were free to relax.

\section{Results and discussion}

\begin{figure}[ht]
 \begin{center}
    \includegraphics[scale=0.75, angle=0.]{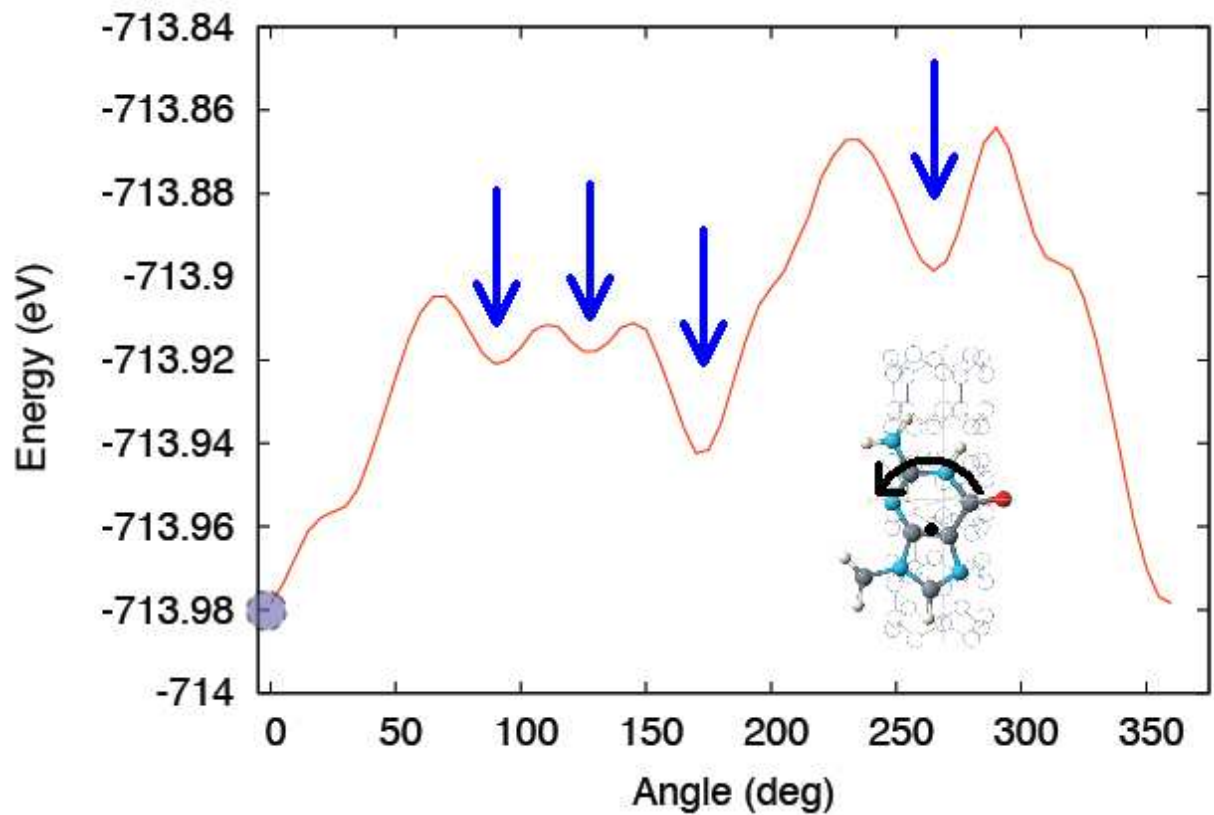}
    \caption{Rotational energy scan for guanine on top of a (5,0) CNT.
    Zero degree orientation (corresponding to global minimum) and rotating direction shown in the inset.
    The blue arrows indicate local minima for specific rotation angles.}
    \label{ROTATION}
  \end{center}
\end{figure}

The initial step in the constrained optimization process resulted in
a configuration of all five nucleic acid bases in which their planes
are oriented almost exactly parallel to the CNT surface. The inset
in Fig. \ref{ROTATION} shows the configuration referred to as the
axial configuration. The base molecule-CNT separation was about 3.2
\AA, which is a little less than the characteristic distance for
$\pi$--$\pi$ stacked systems \cite{pi-stacking}.  The latter does
however strictly apply only for planar entities, the high-curvature
surface of a tube such as (5,0) allows for the $\pi$-orbitals of the
nucleic acid base to come closer before the repulsive interaction
sets in.

The base was translated both along the CNT axis and around its
circumference respectively, maintaining a constant separation of
approximately 3.2 \AA\ from the CNT surface, as determined in the
previous step. The translational scan of the energy surface, as can
be seen from Fig. \ref{PES}, gives an energy barrier of about 0.07
eV for all five molecules. At room temperature, this barrier is
sufficiently large to affect the mobility of the base molecules
physisorbed on the CNT surface and to constrict their movement to
certain directions. The base was then rotated 360$^\circ$, in the
minimum total energy configuration obtained from the previous step.
We found energy barriers of up to 0.12 eV (Fig. \ref{ROTATION}),
resulting in severe hindrance of changes in the orientation of the
physisorbed nucleic acid base. Interestingly, local minima were
found for special rotations corresponding to 90$^\circ$,
120$^\circ$, 180$^\circ$, and 270$^\circ$ (Fig. \ref{ROTATION}).

We emphasize here that for all five base molecules, the calculated
equilibrium configuration was characterized by a separation between
base and CNT surface that was equal to the optimum height chosen in
the previous lateral potential energy surface scan.

\begin{figure}[ht]
 \begin{center}
    \includegraphics[scale=0.50, angle=0.]{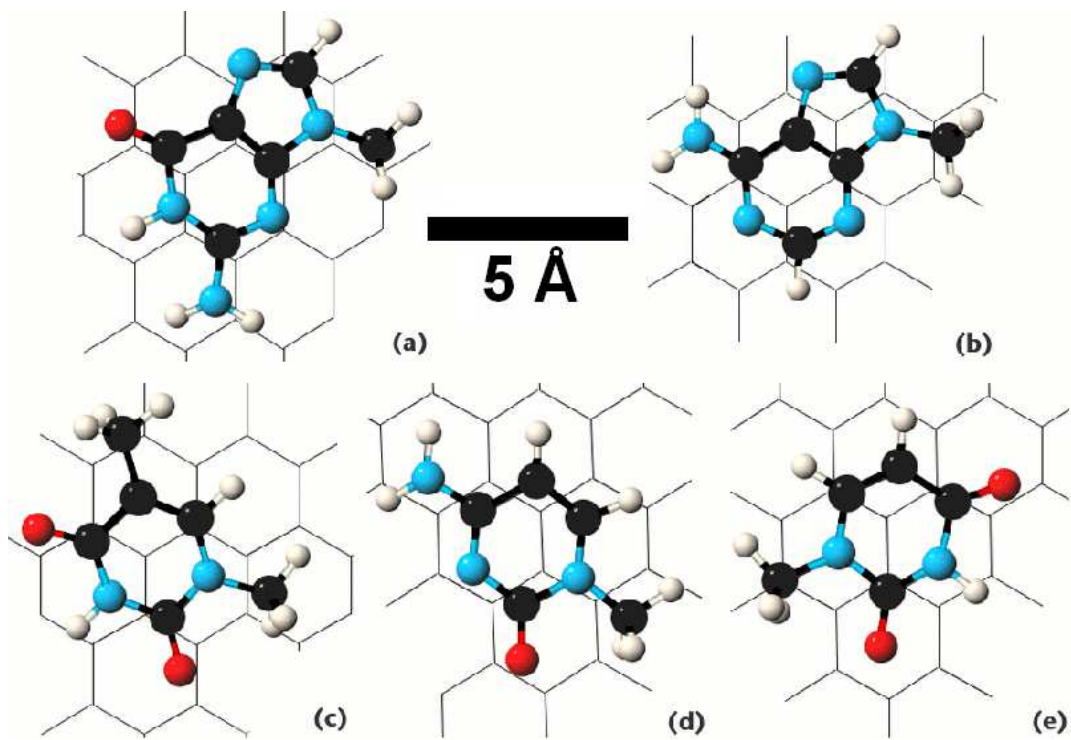}
     \caption{Equilibrium geometry of nucleic acid bases on top of CNT:
     (a) guanine, (b) adenine, (c) thymine, (d) cytosine, and (e) uracil.
     The CNT surface has been flattened into the x-y-plane for clearer visibility.
     Bar indicates scale in figure.}
    \label{EQGEOMETRY}
  \end{center}
\end{figure}

In their equilibrium configuration, the base molecules A, T and U
tend to position themselves on the CNT in a configuration
reminiscent of the Bernal's AB stacking of two adjacent graphene
layers in graphite (Fig. \ref{EQGEOMETRY}). The base molecules G and
C, on the other hand, show a lesser degree of resemblance to the AB
stacking. The interatomic structure of the nucleic acid bases in
their equilibrium configurations underwent virtually no changes when
compared to the corresponding gas-phase geometries, as it could be
expected for a weakly interacting system.

The tendency of the $\pi$--orbitals of the bases and the
graphene-like surface of a CNT to minimize their overlap, in order
to lower the repulsive interaction, helps us understand the observed
stacking arrangement (Fig. \ref{EQGEOMETRY}). The geometry deviates
from the perfect AB base-stacking as, unlike graphene, the six- and
five-membered rings of the bases possess a heterogeneous electronic
structure due to the presence of both nitrogen and carbon in the
ring systems. Additionally, there exist different side groups
containing CH$_3$, NH$_2$, or O, all of which contribute to the
deviation from the perfect AB base-stacking as well.

The binding energy of the system consisting of the nucleic acid base
and the CNT is taken as the energy of the equilibrium configuration
with reference to the asymptotic limit obtained by varying the
distance between the base and the CNT surface in the direction
perpendicular to both the tube axis and the plane of the base
molecule (Table \ref{BE_pol}). G is found to bind most strongly,
while the binding for U with the CNT surface is the weakest.

Table \ref{BE_pol} also includes the polarizabilities of the base
molecules calculated using the Hartree--Fock approach coupled with
second--order M{\o}ller--Plesset perturbation theory (MP2) as
implemented in the {\sc gaussian 03} suite of programs
\cite{GAUSSIAN}. The polarizability of the base molecule
\cite{Seela:2005}, which represents the deformability of the
electronic charge distribution, is known to arise from the regions
associated with the aromatic rings, lone pairs of nitrogen and
oxygen atoms. The calculated polarizability for the purine base G
thus has the largest value, whereas the pyrimidine base U has the
smallest value of polarizability among the five bases.

The CNT-molecule binding energies and the molecular polarizabilities
of the base molecules calculated using MP2 (Table \ref{BE_pol}) show
a remarkable correlation. The polarizability of a nucleic acid base
plays a key role in governing the strength of interaction with the
CNT. This is an expected behavior for a system that draws its
stabilization from vdW dispersion forces, since the vdW energy is
proportional to the polarizabilities of the interacting entities.
The observed correlation thus strongly suggests that vdW interaction
is indeed the dominant source of attraction between the CNT and the
nucleic acid bases.

\begin{figure}[ht]
 \begin{center}
    \includegraphics[scale=0.75, angle=0.]{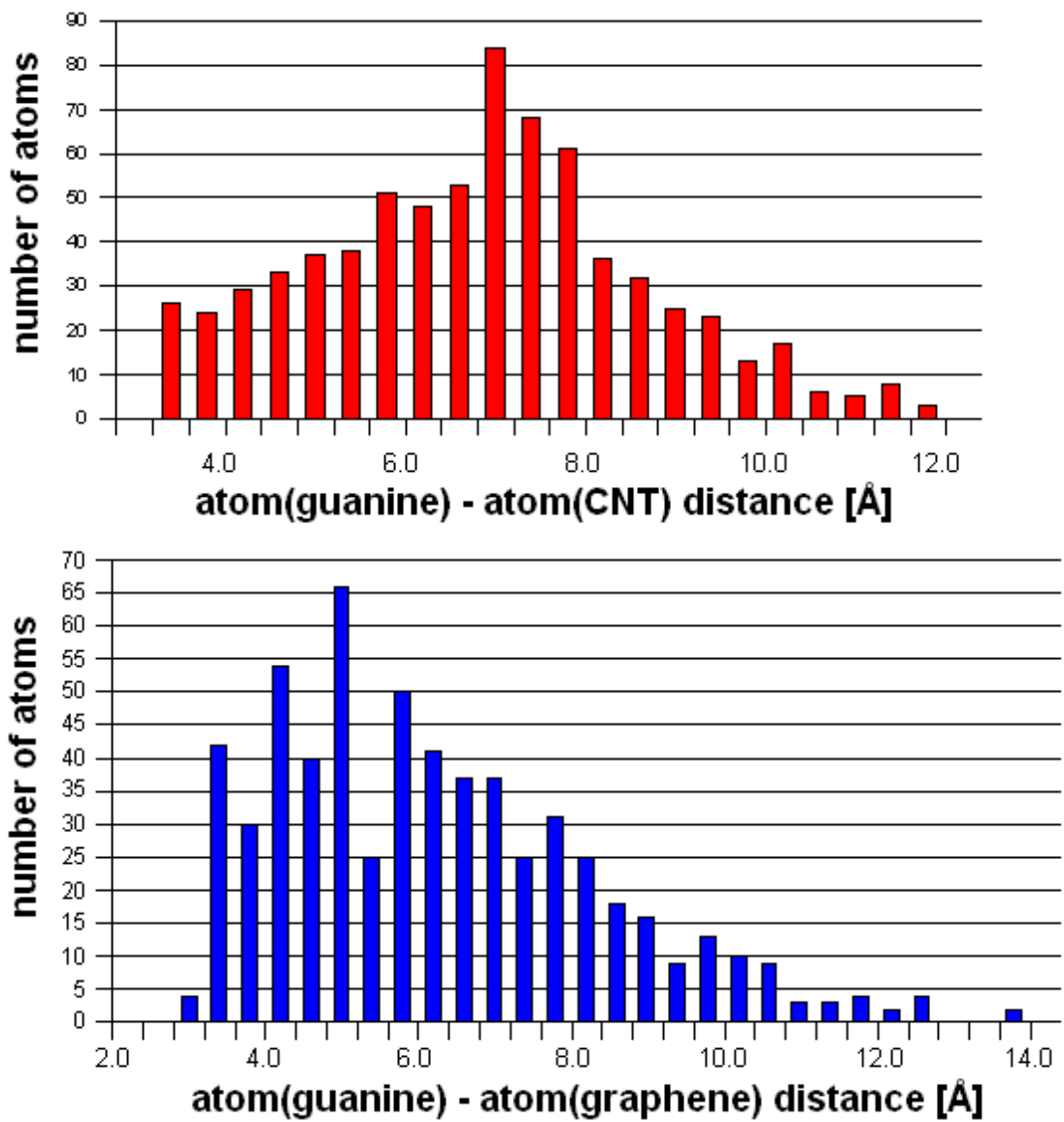}
    \caption{Distance distribution of atoms in guanine relative to carbon atoms
    in (top) CNT and (bottom) graphene.}
    \label{HISTOGRAM}
  \end{center}
\end{figure}

Comparing the present results with those obtained for graphene
\cite{Gowtham:2007}, we clearly see (Table \ref{BE_pol}) that the
binding energy of the base molecules is substantially reduced for
physisorption on small-diameter CNTs with high curvature. While the
curvature allows the nucleic acid base to approach the surface more
closely, the majority of the carbon atoms in CNT are actually
further removed from the atoms of the bases than in the
corresponding case on a graphene sheet (Fig. \ref{HISTOGRAM}). Since
the interaction is clearly dominated by vdW dispersion which falls
off as $1/{r^6}$ with the distance $r$, the overall interaction is
reduced in the case of CNT.

\begin{figure}[ht]
 \begin{center}
    \includegraphics[scale=0.50, angle=0.]{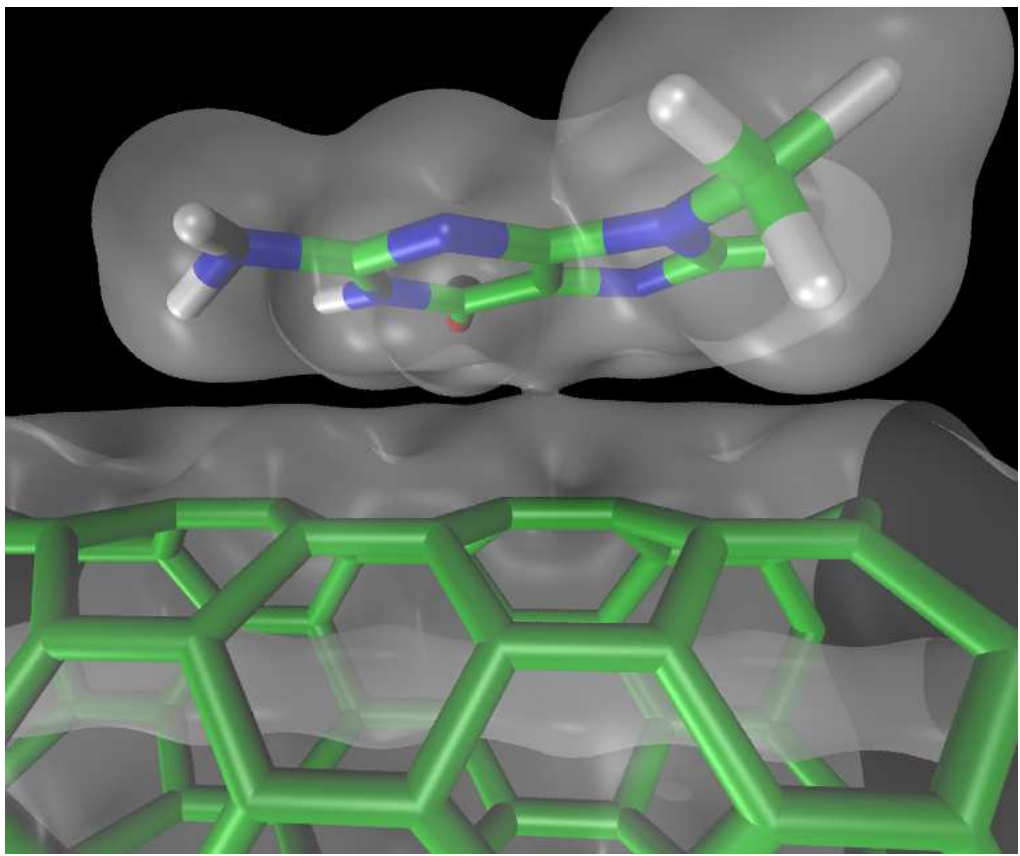}
    \caption{Charge density plot for guanine physisorbed on a (5,0) CNT.
    A small funnel is noticeably connecting the two entities near the C-C bond
    of guanine where the six- and five-membered rings join in the molecule.}
    \label{CHARGE}
  \end{center}
\end{figure}

We furthermore calculated the charge transfer between the bases and
the CNT. For G, we find from the Bader analysis that the CNT
possesses an excess charge of $-$0.08~$e$ and correspondingly a
slight depletion of electrons on G by +0.08~$e$. For A with CNT,
$-$0.05~$e$ were found to have been transferred from the nucleic
acid base to the CNT. These results should be compared with our
corresponding findings from the interaction of nucleic acid bases
with a flat graphene sheet \cite{Gowtham:2007}, where merely
0.02~$e$ were transferred in the case of G. Thus, the higher
curvature of the (5,0) CNT leads to an increased electronegativitiy
which manifests itself in the larger amount of charge transferred to
it. The different behavior of G and A becomes understandable when
one considers that G has a smaller ionization potential than A, and
it is thus easier to remove an electron from G than from A. While
there are no ``whole elementary charges'' transferred in this case,
but only fractions, it still shows that the CNT is able to get more
charge from G than from A. It appears that the charge transfer
originates primarily from the C--C bond that joins the six- and
five-membered ring (Fig. \ref{CHARGE}).

\begin{figure}[ht]
 \begin{center}
    \includegraphics[scale=0.50, angle=0.]{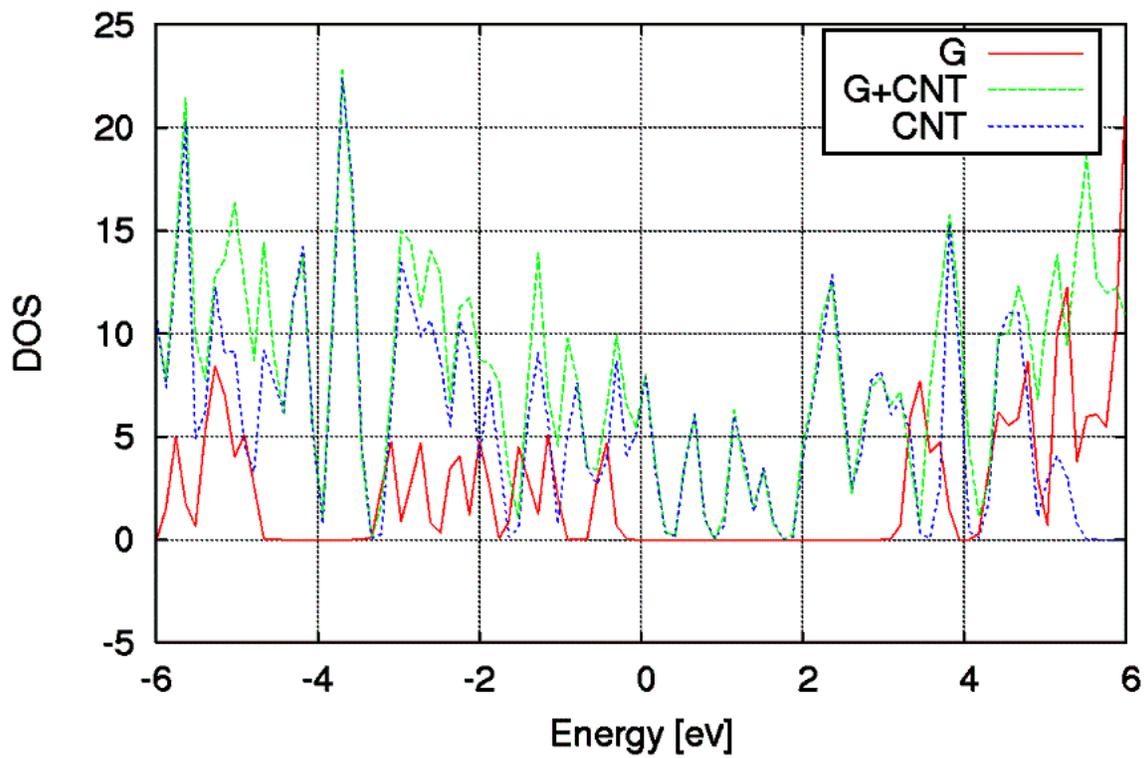}
    \caption{Comparison between the density of states for an isolated guanine
    molecule (G), an isolated (5,0) carbon nanotube (CNT), and the combination
    of the two at equilibrium geometry (G+CNT). }
    \label{DOS}
  \end{center}
\end{figure}

Finally, we also analyzed the density of states (DOS) for the
combined system of base+CNT and compared with the corresponding DOS
for the individual parts, i.e., CNT and nucleic acid base separated
(Fig. \ref{DOS}). We find that the DOS of the combined system is
almost exactly the superposition of the DOS of the individual parts,
in agreement with a recent tight-binding study of DNA-wrapped CNTs
\cite{Seifert:2007}. This finding highlights that the nucleic acid
bases and CNT are interacting rather weakly, and that no significant
hybridization between the respective orbitals of the two entities
takes place.

\section{Conclusions}

In summary, we have investigated the interaction of the five DNA/RNA
base molecules with a (5,0) zigzag CNT of very high curvature by
{\it first-principles} methods. From the calculations, the five
nucleic acid bases are found to exhibit significantly different
interaction strengths with the CNT. Molecular polarizability of the
base molecules is found to play the dominant role in the interaction
strength of the base molecules with CNT. This observation should be
of importance in understanding the sequence-dependent interaction of
DNA with CNTs observed in experiments
\cite{Zheng:2003b,Johnson:2005}.

When comparing the results obtained here for physisorption on the
small diameter CNT considered with those from the previous study on
graphene \cite{Gowtham:2007}, we see that the interaction strength
of nucleic acid bases is smaller for the tube. Thus, it appears that
introducing surface curvature reduces the binding energy between the
base molecule and the substrate. The binding energies for the two
extreme cases of negligible curvature (flat graphene sheet) and of
very high curvature (the (5,0) CNT studied here) represent the upper
and lower boundaries, and it is expected that the binding energy of
bases for CNTs of intermediate curvature is likely to lie in between
these two extremes. Based on the results obtained up to this point,
the hierarchy of the binding energies of the nucleic acid bases to
the graphene-like surfaces of CNTs appears to be universally valid,
as long as the interaction is dominated by vdW forces. Further
studies are currently in progress to consider the effect that
different chiralities may have on the interaction of nucleic acid
bases with high-curvature CNTs.

\section*{Acknowledgments}

The authors acknowledge the availability of computing time from the
National Center for Supercomputing at the University of Illinois at
Urbana-Champaign and thank Dr. Takeru Okada and Prof. Rikizo
Hatakeyama of Tohoku University for helpful discussions. S.G.,
R.H.S., and R.P. would like to thank DARPA for funding. R.H.S. also
acknowledges support from EXC!TiNG (EU Research and Training
Network) under contract HPRN-CT-2002-00317. The research reported in
this document was performed in connection with contract
DAAD17-03-C-0115 with the U.S. Army Research Laboratory.

\newpage

\begin{table}
\begin{center}
\begin{tabular}{cccc}
\hline \hline \multicolumn{1}{c} {base} & \multicolumn{1}{c}
{~~$E_b^{CNT}$ (eV)} & \multicolumn{1}{c} {~~$E_b^{graphene}$ (eV)}
&
\multicolumn{1}{c} {~~$\alpha$ ($e^2a_0^2E_h^{-1}$)} \\
\hline
G  & 0.49  & 1.07 & 131.2 \\
A  & 0.39  & 0.94 & 123.7 \\
T  & 0.34  & 0.83 & 111.4 \\
C  & 0.29  & 0.80 & 108.5 \\
U  & 0.28  & 0.74 & 97.6 \\
\hline \hline
\end{tabular}
\end{center}
\caption{Binding energy $E_b$ of the DNA/RNA nucleic acid bases with
a (5,0) CNT and with a flat graphene sheet as calculated within LDA.
A close correlation with the nucleic acid bases' polarizabilities
$\alpha$ from MP2 calculations can be seen.} \label{BE_pol}
\end{table}

\newpage

\newpage

\newpage

\newpage

\newpage

\newpage

\end{document}